# The Development and Study of High-Position Resolution (50 μm) RPCs for Imaging X-rays and UV photons

C. Iacobaeus[1], P. Fonte[2], T. Francke[3,4], J. Ostling[5], V. Peskov[3], J.Rantanen[4], I. Rodionov[6]

[1] *Karolinska Institute, Stockholm, Sweden*
[2] *LIP and ISEC, Coimbra, Portugal*
[3] *Royal Institute of Technology, Stockholm, Sweden*
[4] *XCounter AB, Danderyd, Sweden*
[5] *Stockholm University, Stockholm, Sweden*
[6] *Reagent Research and Development Center, Moscow, Russia*

# The Development and Study of High-Position Resolution (50 μm) RPCs for Imaging X-rays and UV photons

C. Iacobaeus[1], P. Fonte[2], T. Francke[3,4], J. Ostling[5], V. Peskov[3], J.Rantanen[4], I. Rodionov[6]

[1] *Karolinska Institute, Stockholm, Sweden*

[2] *LIP and ISEC, Coimbra, Portugal*

[3] *Royal Institute of Technology, Stockholm, Sweden*

[4] *XCounter AB, Danderyd, Sweden*

[5] *Stockholm University, Stockholm, Sweden*

[6] *Reagent Research and Development Center, Moscow, Russia*

**Abstract**

Nowadays, commonly used Resistive Plate Chambers (RPCs) have counting rate capabilities of ~$10^4$Hz/cm$^2$ and position resolutions of ~1cm. We have developed small prototypes of RPCs (5x5 and 10x10cm$^2$) having rate capabilities of up to $10^7$Hz/cm2 and position resolutions of 50 μm ("on line" without application of any treatment method like "center of gravity"). The breakthrough in achieving extraordinary rate and position resolutions was only possible after solving several serious problems: RPC cleaning and assembling technology, aging, spurious pulses and afterpulses, discharges in the amplification gap and along the spacers. High-rate, high-position resolution RPCs can find a wide range of applications in many different fields, for example in medical imaging. RPCs with the cathodes coated by CsI photosensitive layer can detect ultraviolet photons with a position resolution that is better than ~30 μm. Such detectors can also be used in many applications, for example in the focal plane of high resolution vacuum spectrographs or as image scanners. © 2001 Elsevier Science. All rights reserved

*Keywords*: RPC, X-ray imaging, UV detection

## 1. Introduction

At present, commonly used RPCs typically have rate capabilities of ~$10^4$Hz/cm2 and position resolutions of ~1cm [1]. In [2] we have demonstrated that a very high position resolution (~50 μm) could be achieved with small gap (0.1-0.5 mm) RPCs, the cathode of which was coated by the porous CsI secondary electron emitter. It is remarkable that this excellent accuracy in position measurements was obtained in a simple counting (digital) mode without using any analogous interpolation method (like "center of gravity"). In a subsequent work [3] we showed that the counting rate characteristics of the RPCs could be considerably improved by using low resistivity materials, which at the same time still enabled the RPCs to be protected from sparks. In [4] an attempt was made to unify both approaches and develop an RPC with a CsI secondary electron converter to be used for tracking. The aim of this work is to investigate the possibility of building high rate high-position resolution RPCs for imaging x-rays and UV photons.

## II. Experimental set up

Our experimental set up is schematically presented in Fig.1. It basically consists of a test gas chamber, inside which various designs of RPCs could be installed, and a radiation source (an x-ray tube or a Hg lamp). The test chamber had three windows: one transparent for X-rays (6-60 keV), another one for UV (~192 nm) photons and the third one for visible light. Many different designs of RPCs were tested; all had sizes of 5x5 cm$^2$ and 10x10 cm$^2$ with gaps between the electrodes of 0.4 mm or 0.3 mm. The main purpose of these studies was to evaluate various commercially available low resistivity materials for the RPC electrodes and designing optimum spacers and grooves around the spacers. Some of our preliminary results were presented in [5]. In this paper we will focus only on the final designs of the RPCs. Their cathodes were manufactured from GaAs ($\rho\sim10^6$-$10^8\Omega$cm) or Si ($\rho\sim10^4\Omega$cm) wafers. The anodes were manufactured from Pyrex ($\rho\sim10^{14}\Omega$cm) or Pestov glass ($\rho\sim10^{10}$ $\Omega$cm). The anodes were covered with Cr strips of 30 or 50 μm pitches. In most designs 20 strips in the central region were connected to individual charge-sensitive amplifiers, whereas the others were connected in groups. A prototype oriented on medical imaging applications had an ASIC readout of all (~100) the individual strips. To monitor possible discharges inside the volume of the RPC and along the spacers a PMT was used attached to the test chamber's window transparent to visible light. For complimentary visual observations a mirror-based optical system was used. In earlier tests [5] we found that the performance of the RPC depended very much on the quality of the surface of the electrodes. This is why prior to assembling, all electrodes were carefully cleaned: first in various solvents, then washed ultrasonically in soapy water and afterwards, in clean distilled and deionized water. To remove the oxidation layer, the cathodes were also additionally etched by the plasma discharge in the Ar atmosphere. The assembling of the detector was done in a clean room (class of 100).

For comparative studies we also used various metallic parallel-plate avalanche chambers (PPACs). Their electrodes were made from mirror polished metallic plates of 5x5 cm$^2$. The gap between the electrodes was 0.4 mm. Some designs had glass electrodes with the inner surfaces coated by a vacuum evaporated Al layer. For position resolution measurements we used the glass plated with the Al strips with a 50 μm pitch. As an X-ray source, a Kevex tube with exchangeable Fe, Mo or W anodes was used. In contrast to our previous developments [4], the RPCs for X-ray imaging did not have an CsI converter. The collimated x-ray beam (50x50 μm$^2$) was introduced near the cathode and parallel to it. The collimator could be moved in a direction perpendicular to the anode strip with a micron accuracy. The test chamber was attached to a specially designed table allowing a 3-D alignment with a micron accuracy. For obtaining a 2-D X-ray image, the object (for example a test phantom) was moved between the detector and the X-ray tube in a direction perpendicular to the detector's electrodes. In these measurements the x-rays were collimated with a slit of 50 μm in width, placed close to the RPC cathode and oriented parallel to the electrodes and perpendicular to the readout strips. For measurements with the UV, the cathodes of the RPCs were coated with a 0.4 nm thick CsI layer. For the position resolution, a screen with a slit of 50 μm in width oriented perpendicular to the electrode was used. The screen with the slit could be moved in a direction perpendicular to the anode strips with a micron accuracy. The detectors were tested in various Ar, Xe and Kr based mixtures at pressures of 1-3 atm.

## III. Results
### III-1. X-ray detectors

The main challenge in developing high rate RPCs for X-ray imaging is that in contrast to the tracking device [4] there is no external trigger. As a consequence of that, the noise (or spurious) pulses and micro breakdowns which could be almost "invisible" in the case of the triggered detector, may create serious problems in image reconstruction.

Thus the requirements of the detector's performance for imaging applications are much higher than for the case of the tracking device.

### a) Noise pulses and afterpulses

As was mentioned above, the key in achieving a good performance of the detectors was the quality of the electrode's surface. If their surface contained any depositions (thin dielectric films or microparticles) or an oxidation layer, then both the RPC and the PPAC suffered from the noise pulses. As an illustration Fig. 2 shows a pulse height spectra of noise pulses and, for comparison, a pulse height spectra of single electrons (produced by UV radiation) for the RPC with the cathode made of Si exposed to air and which thus had a thin oxidized layer. The measurements were done at gas gains of ~$10^4$. One can see that the mean amplitude of the noise pulses is considerably larger than the mean of the single electron spectra. Note, that as follows from our measurements, at gains of ~$10^4$ the micro gap RPC operated in a limited proportional mode so the number of primary electrons triggering the noise pulse was in reality larger than could be seen from a simple comparison of the pulse height spectra. The nature of this noise pulse is associated with the jets of electrons emitted from these dielectric insertions and layers (see [6]). The counting rate of the noise pulses increased sharply with the applied voltage and with the intensity of the x-ray or UV radiation. If the intensity of the external radiation suddenly decreased, the rate of the noise pulses decreased exponentially with a time constant, which could vary depending on the conditions from 0.1 sec. up to 20 min. (see Fig. 8 in [6]). We call this phenomenon "noise pulses decaying effect". There are two main negative effects associated with the noise pulses: 1) due to the statistic fluctuations, the amplitude of some noise pulses could be large enough to provoke breakdowns [6] and thus lower the maximum achievable gains of the detectors; 2) the "noise pulses decaying effect" spoils the quality of the scanned high-contrast images and images that change during acquisition. The noise pulses also produce bright concentrated spots and clusters of spots in images (see Fig.15 in [5]), which may be confusing and unacceptable for some medical applications. Thus only RPCs assembled from well cleaned and oxidation free Si and GaAs can be used in high rate applications. The other sources of noise pulses are streamers and microbreakdowns along the spacers between the electrodes. Even spacers with well developed surfaces and discharge protecting grooves around them (which normally perform well at low rates [5]) become a source of serious trouble under the strong x-ray's radiation. This is why for high-rate applications the best solution is to simply shield the spacers against the direct hit of the external radiation. Below we will present the results obtained only with detectors with shielded spacers.

### b) Aging

Besides the noise pulses, the second serious problem we came across in developing high rate RPCs was aging. In most gases, after the detector operated for some time at counting rates of $10^5$Hz/mm$^2$, depositions on electrodes appeared. They caused not only noise pulses, but also an unstable detector response with time and a so called "memory effect" [6]. After many experiments we found several gas mixtures which did not have any signs of aging and at the same time allowed the detectors to operate at gains that were high enough to give good position resolutions. An example could be various Xe and Kr mixtures with ~20% of $CO_2$ used as a quencher. However, even in these gas mixtures, any tiny impurity (for example vapors of some glues and plastic materials) could cause a dramatic aging effect. In addition to this, some materials could decompose under direct X-ray or VUV radiation and, thus, also trigger an aging effect. This is why it was absolutely necessary to shield against the direct irradiation of not only spacers, but also the detector's constructing parts (supporting frames, wires ect.). Only after all of these precautions were we able to achieve high-rate, high-position resolutions with our RPC and PPAC detectors (see below).

### c) Rate characteristics

In Fig. 3 typical gain vs. rate characteristics are presented for the RPC (the cathode of which was made of GaAs ρ ~$10^5$Ωcm) and for the PPAC for comparison. The last point of each gain vs. rate curve for the PPAC corresponds to the beginning of breakdowns. One can see that with the low resistivity RPC one can reach gains and counting rates as high as with the metallic PPAC. There is however, an essential difference: at the discharge points powerful (and thus destructive) sparks appear in the PPAC, whereas in the case of the RPC-there were no sparks at all, only mild "streamers". Note that at some range of the RPC's cathode resistivity, a glow discharge may appear instead of the "streamers" [4,5] which may also be destructive for the electronics as well as for the detector's electrodes. As was shown in the [4], glow discharges could be suppressed by increasing the quenching concentration to 30-40%.

**d) Position resolution**

Position resolution measurements were performed with a standard line- pair phantom (a phantom with 7 and 10 lp/mm) and also with a 50x50 μm collimator moving perpendicular to the strips (see section II). Results for the one-dimensional measurements of the phantom image performed with the RPC described above are presented in Fig. 4. The maximum counting rate during this image taking was $10^5$Hz/strips, mean photon energy ~20 keV. One can see that lines with a 70 μm pitch were easily resolved. Results obtained with a 50x50 μm$^2$ collimator were very similar to that published previously (see Fig. 19 in [5]) and confirm that a position resolution of better than 50 μm could be achieved in a simple counting mode, without applying any analogous interpolation method. The same space resolution was obtained with the PPAC, however at high counting rates they had occasional sparking, which certainly limits their industrial or medical applications.

**III.2 Results with the UV detectors**

These measurements of the detector's position resolution were performed with a moving slit. When the slit was aligned exactly along one strip, the maximum counting rate was measured from this particular strip (see Fig. 5). When the slit was aligned in between two strips, the highest counting rates were measured from these strips. This picture was periodically repeated during the slit movement. One can conclude from this data that a position resolution of about 30 μm was achieved in a digital mode of measurement. The RPC could operate at counting rates of up to $10^4$-$10^5$ Hz/μm$^2$ without any sparking or charging up effect. The quantum efficiency of this detector depended on the angle φ (between the beam and the surface) at which the Hg light hit the CsI cathode. For a 192nm light, the quantum efficiency was 15% at φ=30° and 2.7% at φ=5°.

**IV. Discussion and Conclusion**

The breakthrough in achieving extraordinary rate and position resolutions was only possible after solving several serious problems: the choice of optimum materials and spacer design, RPC cleaning and assembling technology, aging, spurious pulses, discharges in the amplification gap and along the spacers. We demonstrated that the PPAC and the RPC have the same position resolution and rate characteristics, however the RPCs are spark-protected, which opens a possibility of their wide range of applications, for example in X-ray and UV imaging. Based on the results of this work we have developed a modified version of the RPC [7] for medical applications. Some images obtained with this detector in scanning mode can be found in [7,8]. Since these detectors operate in a photon counting mode, the dose delivered to a patient is reduced by a factor of 5-10 compared to usual films. We are now working on applications of the photosensitive RPC for scanning UV images and as a linear detector in a focal plane of vacuum spectrographs.

**References:**

[1] R Santonico, "RPCs challenge for next two years" Submitted to Nucl. Instr. in Phys. Research
[2] P.Fonte et al., Nucl. Instr. in Phys. Research A454 (2000) 260
[3] P. Fonte et al., Nucl. Instr. in Phys. Research A431 (1999) 154
[4] T. Francke et al., Preprint Physics/0206039, Jun. 2002
[5] P. Carlson et al., Preprint CERN, CERN-OPEN-2001-076, November 2001
[6] C. Iacobaeus et al., Preprint Physics /0112013, Dec. 2001
[7] T. Francke et al., Nucl. Instr. in Phys. Research A471 (2001) 85
[8] www.xcounter.se

6    Computer Physics Communications

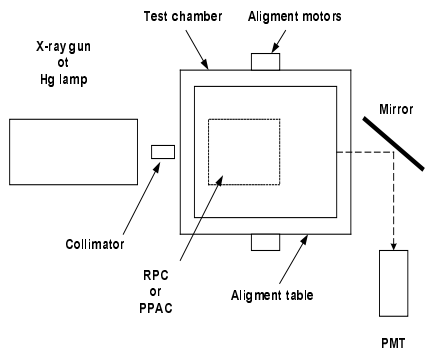

Fig.1. A schematic drawing of the experimental set up

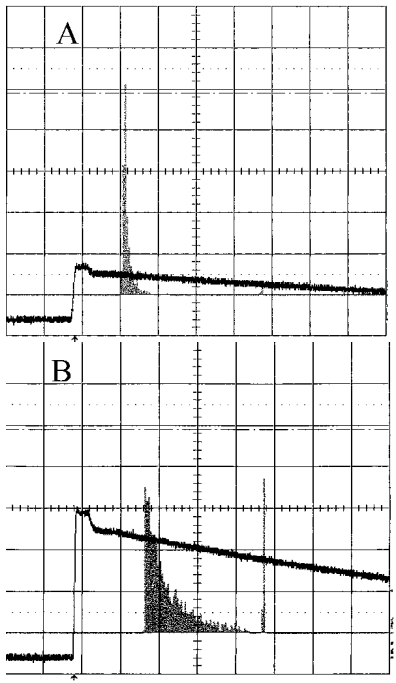

Fig. 2. Pulse height spectra of signals from RPC measured in the case of single primary electrons produced from the cathode by UV emission (a) and in the case of noise pulses (b). Gas mixture Xe(40%)+Kr(40%)+CO$_2$ (20%) at p=1atm

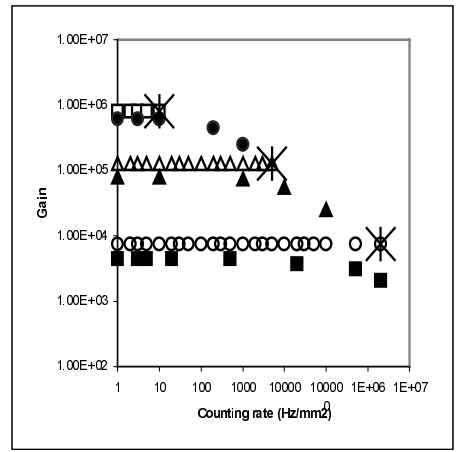

Fig. 3. Gain vs. rate for PPAC (open symbols) and for GaAs cathode (filled symbols) in Xe+Kr+20%CO$_2$ with 6 KeV X-rays. "Star" symbols indicate beginning of breakdowns in the PPAC.

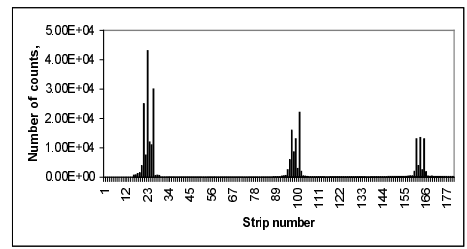

Fig.4. One dimensional digital image of the phantom with 5.5, 6 and 7 lp/mm. Strip pitch 50 μm. The acquisition time 0.1 sec. X-rays were produced from a Mo target. Gas mixture Xe(40%)+Kr(40%)+CO$_2$ (20%) at p=1atm

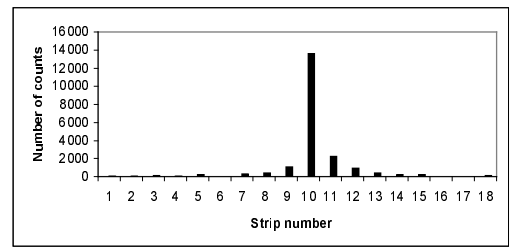

Fig.5 Counting rate measured from various strips when the slit is illuminated by UV aligned along strip # 10. Strip pitch was 30 μm